\documentclass[11pt,twoside]{article}

\usepackage{asp2006}
\usepackage{epsf}
\usepackage{psfig}
\usepackage{lscape}

\begin{document}
\title{Probing the Impact of Stellar Duplicity on Planet Occurrence}   
\author{A.~Eggenberger,$^{1,2}$ S.~Udry,$^{2}$
G.~Chauvin,$^{1}$ J.-L.~Beuzit,$^{1}$
A.-M.~Lagrange,$^{1}$ and M.~Mayor$^{2}$}   
\affil{$^1$Laboratoire d'Astrophysique de Grenoble, Universit\'e Joseph
Fourier, BP 53, 38041 Grenoble Cedex 9, France}    
\affil{$^2$Observatoire de Gen\`eve, Universit\'e de Gen\`eve, 51 ch. des
Maillettes, 1290 Sauverny, Switzerland}    

\begin{abstract} 
The presence of a stellar companion closer than $\sim$100 AU is likely to affect
planet formation and evolution. Yet, the precise effects and their actual impact
on planet occurrence are still debated. To bring observational constraints, we
have conducted with VLT/NACO a systematic adaptive optics survey for close
stellar companions to 130 solar-type stars with and without planets. 
In this paper we present observational and preliminary statistical results from
this survey. Observational results reveal about 20 true companions,
of which 4 are new companions to planet-host stars. As to preliminary 
statistical results, they suggest that circumstellar giant planets are less 
frequent in binaries closer than $\sim$100 AU than around single stars, in 
possible agreement with the theoretical studies that predict a negative impact 
of stellar duplicity on giant planet formation in binaries closer than 
$\sim$100 AU. These statistical results will need confirmation, however, as they
are severely limited by small sample sizes. 
\end{abstract}

\section{Introduction} 
\label{intro}

Nearby G-K dwarfs, which are first-choice targets for Doppler planet 
searches, are more often found in binaries or multiple systems than in 
isolation \citep[e.g.][]{Duquennoy91,Halbwachs03,Eggenberger04}. 
This observation, coupled with the growing evidence that many young 
binaries possess circumstellar or circumbinary disks susceptible of
sustaining planet formation \citep[][and references therein]{Monin07}, raises the question of the existence and frequency
of planets in and around star systems of different types. 

Over the past years, binaries have become increasingly 
interesting targets of planet searches and studies. Doppler surveys have shown 
that circumstellar giant planets exist not only in wide binaries, but 
also in the much closer spectroscopic binaries, implying that planets
may be quite common in various types of binaries and hierarchical triples. 
Nonetheless, double stars closer than 2-6\arcsec\ present technical difficulties 
for Doppler planet searches (light contamination from the companions; 
\citet{EggenbergerUdry07}). Most Doppler surveys avoid these systems 
accordingly, and the actual frequency of 
planets in binaries closer than $\sim$200 AU is as yet unknown. Similarly, 
the existence of lower mass ($<$0.2 M$_{\rm Jup}$) circumstellar planets  
and the existence of circumbinary planets are essentially unconstrained from 
the present observations. As a consequence, we will only consider
circumstellar giant planets in this paper.

Theoretical studies have shown that the most 
sensitive issue regarding the occurrence of giant planets in binaries closer
than $\sim$100 AU seems to be whether or not these planets can form in the first place. 
Within the core accretion paradigm, planetesimal 
accretion can be significantly perturbed by the presence of a stellar 
companion closer than $\sim$100 AU. Yet, these perturbations are usually 
not strong enough to fully halt accretion \citep{Thebault04,Thebault06,Marzari07}. From this point of view, we thus 
expect giant planets to be present, though possibly less frequent, in several
types of binaries closer than $\sim$100 AU. Within the alternative disk 
instability model, the impact of stellar duplicity on disk fragmentation is 
more debated. According to \citet{Nelson00} and \citet{Mayer05},  
disk fragmentation is generally suppressed by the presence of a stellar companion 
within 100 AU, while according to \citet{Boss06} disk fragmentation is 
promoted by the presence of a stellar companion with a periastron less
than $\sim$50 AU. If disk
instability is a viable formation mechanism for giant planets, it is thus not
clear yet whether we should expect more or less giant planets in
intermediate binaries (50-100 AU) than around single stars. Nonetheless, 
the claim by \citet{Mayer05} that stellar duplicity may inhibit giant planet 
formation via disk instability but not via core accretion in binaries separated by 
60-100 AU, is particularly interesting and deserves further investigation. 
Indeed, if true, this statement would imply that planets residing in 60-100 AU binaries 
provide a unique means to probe the main formation mechanism for giant planets. 
In any case, quantifying the occurrence of giant 
planets in binaries closer than 200 AU, and studying how this occurrence 
varies with binary semimajor axis, may provide us with additional 
clues to better understand giant planet formation. 

To quantify the impact of stellar duplicity on planet occurrence in binaries
separated by $\sim$35-250 AU, and to test whether or not the occurrence of 
giant planets is reduced in binaries closer than $\sim$100 AU, we are 
conducting a large-scale adaptive optics search for stellar 
companions to $\sim$200 solar-type stars with and without planets
\citep{Eggenberger07}. To cover a
substantial fraction of the sky, our main program is divided into two 
subprograms: a southern survey (130 stars) carried out with NAOS-CONICA (NACO) on the Very 
Large Telescope (VLT), and a northern survey (about 70 stars) carried out with PUEO on the 
Canada-France-Hawaii Telescope (CFHT). The NACO survey is now almost completed,
while the PUEO survey is still halfway. In this paper we present observational 
and preliminary statistical results from the NACO survey.

\section{The NACO survey}
\label{naco}

\subsection{Overview}

If the presence of a nearby stellar companion 
hinders planet formation or drastically reduces the potential stability zones,
the frequency of planets in binaries closer than a given separation (modulo
eccentricity and mass ratio) should be lower than the nominal frequency of planets around single stars.
Alternatively, if the presence of a nearby stellar companion stimulates planet
formation one way or another, planets should be more common in binaries with a
specific range of separations (again modulo eccentricity and mass ratio) than 
around single stars. Reversing these statements, studying the multiplicity of
planet-host stars relative to that of similar stars but without planetary 
companions, may be a means of quantifying whether or not stellar duplicity  
impacts planet formation and/or evolution.

Since 2002, direct imaging has been 
used by several groups to detect stellar companions close to planet-host stars
\citep{Luhman02,Patience02,Mugrauer05,Mugrauer06,Chauvin06}. 
Yet, to push the investigation a step further and to study the impact of 
stellar duplicity on planet occurrence, one needs not only a sample of 
planet-host stars, but also a control sample of non-planet-bearing stars against 
which to compare the results. Such a control sample is also essential to take 
into account the selection effects against close binaries ($<$2-6\arcsec) 
in Doppler planet searches. The lack of a well-defined
control sample is the major limitation that prevents the above-mentioned surveys
to draw robust conclusions on the impact of stellar duplicity on planet
occurrence. To overcome this limitation and to be as rigorous as possible, we 
included in our NACO survey both a subsample of planet-host stars and a control 
subsample of nearby field stars from our CORALIE planet search program showing  
no obvious evidence for planetary companions from radial-velocity measurements.

\subsection{Sample, Observing Strategy and Observations}

Our NACO survey relies on a sample of 57 planet-host stars, together with 73
control stars. The planet-host star subsample comprises nearby stars: (i) known 
to host a planet from Doppler surveys; (ii) visible from 
Paranal; and (iii) not appearing in previous, deep and relatively wide-field
adaptive optics surveys, to avoid repeating existing observations. 
The control star subsample contains nearby stars: (i) belonging
to the CORALIE planet search sample \citep{Udry00}; 
(ii) with right ascension, declination,
visual magnitude, color, and parallax as close as possible to the corresponding
quantities of one of the planet-host stars; (iii) showing the least possible
radial-velocity variations suggestive of the presence of a stellar or planetary
companion. The larger number of control stars is intentional, as a few  
stars observed in other adaptive optics surveys will be 
added to the planet-host star subsample for the statistical analysis 
(see Sect.~\ref{stat}). 

\begin{figure}[!ht]
\begin{center}
\plottwo{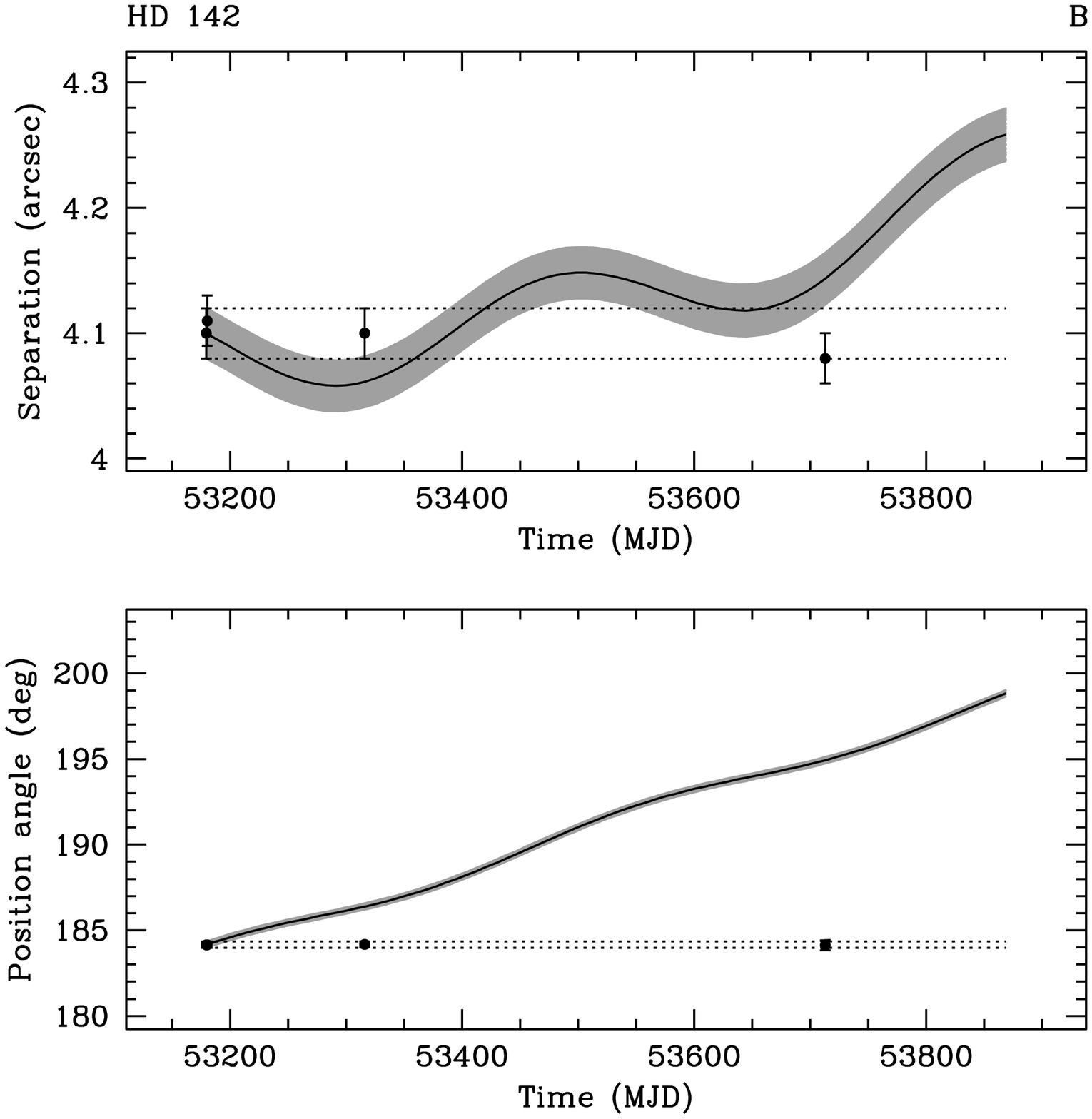}{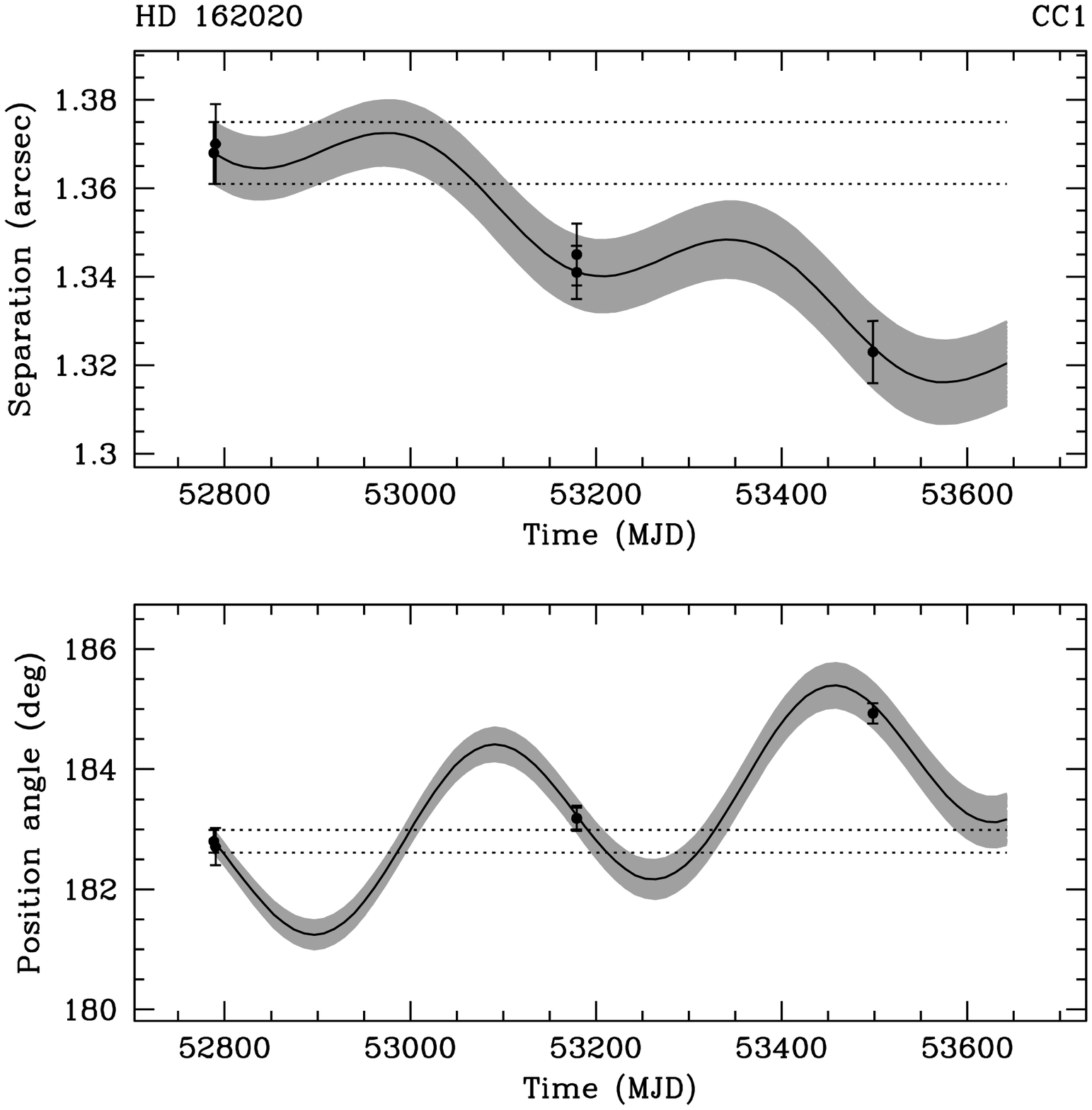}
\end{center}
\caption{Examples of multiepoch astrometry from our NACO survey. Solid lines
depict the evolution of angular separation and position angle for background
objects with negligible proper motion. The gray zones are the related
uncertainties. Dots represent our NACO observations and dotted lines depict the
evolution expected for bound systems without significant orbital motion over the
survey time span. The left panels show an example of true companion, while the
right panels show an example of unrelated background star.}
\label{multi_astro}
\end{figure}

The survey observing strategy consisted of taking a first image of each of our 
targets (planet-host and control stars) to detect companion candidates. To
distinguish true companions from unrelated background  
stars, we relied on two-epoch astrometry. Since most our targets are within 50
pc and have a proper motion above 0.1\arcsec\,yr$^{-1}$, astrometric parameters 
of bound systems are indeed not expected to vary much over a few years, except for 
some orbital motion in the closest systems (Fig.~\ref{multi_astro}, left). 
On the other 
hand, astrometric parameters of background objects without significant proper 
motion should vary according to the proper and parallactic motion of the 
primaries (Fig.~\ref{multi_astro}, right). For relatively wide and bright companion candidates 
($\rho$\,$>$\,10\arcsec, $K$\,$<$\,14), a preexisting astrometric epoch could 
usually be found in the 2MASS catalog \citep{Skrutskie06}, meaning that only one NACO observation 
was needed to identify true companions. However, due to the high angular 
resolution of NACO, 
we could not rely on such preexisting data on a regular basis. To reject from 
the statistics the numerous background stars we thus tried to reobserve the 
targets with companion candidates at a later epoch during the survey.

Our observations were spread over six different runs carried out
between December 2002 and December 2005. For each target, we recorded unsaturated 
images in narrowband filter within the $H$ or $K$ band, the choice depending on 
atmospheric conditions. For an optimal sampling of the point spread 
function, the field of view was 13\arcsec\ for observations in the 
$H$ band and 27\arcsec\ for observations in the $K$ band. 
For targets within 50 pc, the 13\arcsec\ field of view 
translates into a projected separation range of a few AU (diffraction 
limit) to about 325 AU. Recalling the theoretical predictions mentioned in 
Sect.~\ref{intro}, this means that our survey probes a large fraction of the 
separation range where the presence of a stellar companion should 
affect giant planet formation (hence giant planet occurrence) to some 
degree.

\subsection{Observational Results}

Our data revealed 95 companion candidates in the vicinity of 33 targets. On the
basis of two-epoch astrometry, we identified 19
true companions, 2 likely bound objects, and 34 background stars. The 
remaining 40 companion candidates (near 16 targets) either lack second-epoch 
measurements (most objects), or have inconclusive astrometric results due to 
insufficiently sensitive images at one epoch (few objects). 
The low likelihood of chance alignment shows that two of these 
40 objects are very likely bound ($P$\,$<$\,0.1\%), while ensemble statistics 
indicates that a few additional true companions might hide among these candidates.
Follow-up observations are underway to identify these remaining bound systems. 

Among planet-host stars, we discovered two very low mass companions to 
HD\,65216, an early-M companion to HD\,177830, and we resolved the previously 
known companion to HD\,196050 into a close pair of M dwarfs. 
Besides these discoveries, 
our data confirm the bound nature of the companions to HD\,142, HD\,16141, 
and HD\,46375. 
The remaining 11 true companions and the two likely bound objects all orbit 
control stars. These companions are late-K stars or M dwarfs, and have projected
separations between 7 and 505 AU. We refer the reader to \citet{Eggenberger07}
for additional information on all these systems. 

\begin{figure}[!ht]
\begin{center}
\plottwo{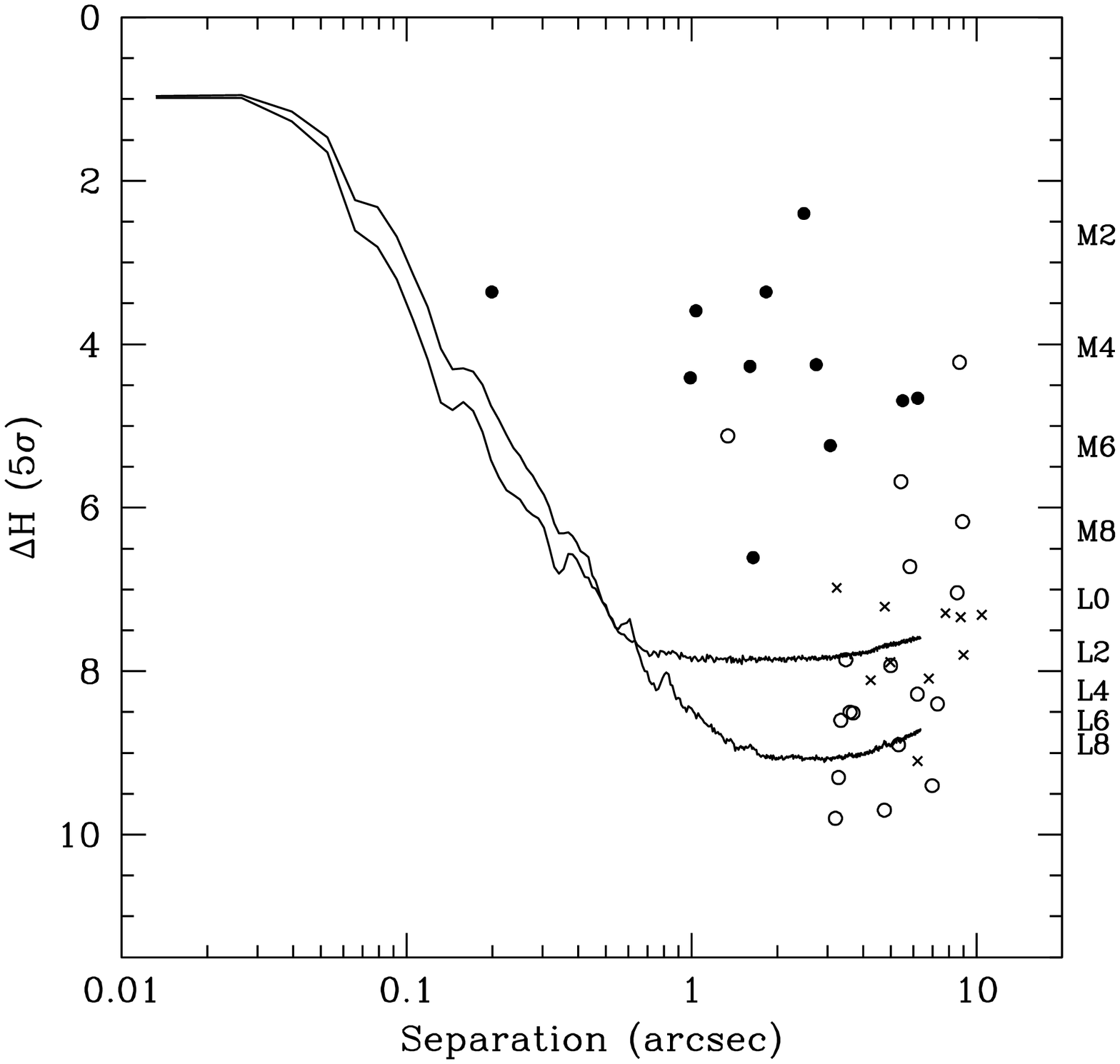}{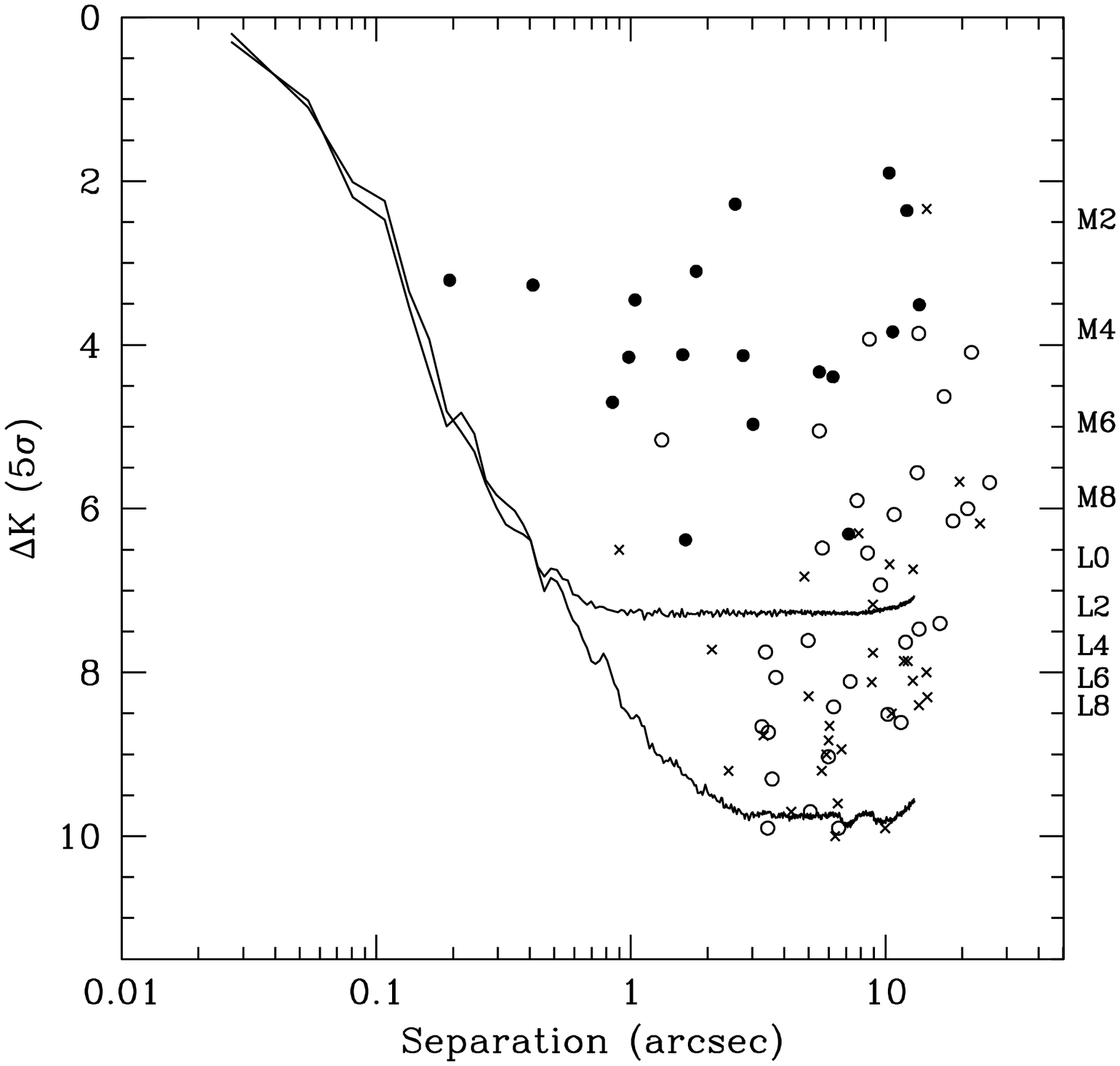}
\end{center}
\caption{Sensitivity limits and detections from our NACO survey. Dots represent
bound and likely bound companions, open circles represent unbound objects, 
and crosses denote companion candidates with only one astrometric
epoch. Solid lines are the median detection limits obtained with the two
different detectors of CONICA (a detector change occurred in the middle of our
survey). Labels on the right-hand side of each plot show the relationship
between magnitude (narrowband photometry) and spectral type for companions to a 
typical old K0 dwarf.}
\label{detlim}
\end{figure}

The typical sensitivity of our survey enabled us to detect stellar companions
down to $\sim$M5 dwarfs at 0.2\arcsec\ and down to the L-dwarf domain above 
0.65\arcsec\ (Fig.~\ref{detlim}), providing us with a very complete census of the stellar 
multiplicity among our 130 targets.  This observational material forms 
an unprecedented data set to study the impact of stellar duplicity on planet 
occurrence.

\section{Preliminary Statistical Analysis}
\label{stat}

Although the inclusion of a well-defined control subsample reduces to their 
lowest value the corrections related to selection effects, some minor 
corrections and refinements in the definition of our two subsamples are still 
needed to properly quantify the impact of stellar duplicity on planet 
occurrence. A first debatable point is the exact definition of the control 
subsample, in particular regarding the non-planet-bearing status of these stars. 
Another issue is the relatively small number of true companions ($<$20) that 
will likely render our statistical results quite sensitive to the
exact definition of the two subsamples. 

To get a first estimate of the impact of stellar duplicity on planet 
occurrence and to test the sensitivity of our results to the exact definition 
of each subsample, we performed a preliminary analysis based on two 
different sample redefinitions: (i) a loose redefinition where both subsamples 
were little modified except for an homogeneous cut-off at close separation
($\sim$0.7\arcsec); 
(ii) a more stringent redefinition where both subsamples
were limited in distance to 50 pc, and where control stars
showing any type of radial-velocity variation were rejected. This additional selection was
aimed at keeping in the control subsample as little potential planet-host stars
as possible. Yet, by being too severe on this point we destroy the homogeneity
in the selection of the control and planet-host subsamples 
and we may bias the results by
rejecting more stars with close stellar companions from the control subsample
than from the planet-host star subsample. Therefore, an optimum has to be found
and the two cases presented here (i and ii) are intended to be some kind of lower 
and upper limits surrounding this optimum. Hereafter, the loosely redefined 
subsamples will be called ``full'' subsamples, while the more refined 
subsamples will be called ``redefined''. 

To both the full
and redefined planet-host subsamples we added the stars observed by 
\citet{Patience02} and \citet{Chauvin06} when their sensitivity limits were 
better than ours in terms of contrast and field of view.
We also assumed that the 40 companion candidates with only one astrometric epoch
were unrelated stars, except for the 
two objects with a low likelihood of chance alignment ($P$\,$<$\,0.1\%). 
Finally, for each filter ($H$ and $K$) we defined 
a complete detection zone by using one of the worst sensitivity limit of the 
survey for separations up to 6.3\arcsec. Angular separations were converted 
into projected separations ($r$), and then
into mean semimajor axes ($a$), using the parallaxes of the primaries and the
statistical relation $a$\,$=$\,1.26\,$r$ from \citet{Fischer92}.

To quantify the global impact of stellar duplicity on giant
planet occurrence in binaries with mean semimajor axes between 35 and 250 AU, we
computed the binary fraction for the four subsamples described above. The binary fraction 
of planet-host stars is $5.5$\,$\pm$\,$2.7$\% (4/73) for the full subsample 
and $4.9$\,$\pm$\,$2.7$\% (3/62) for the redefined subsample. For control 
stars, we obtain binary fractions of $13.7$\,$\pm$\,$4.2$\% (9/66) and 
$17.4$\,$\pm$\,$5.2$\% (9/52) for the full and redefined subsamples, 
respectively. These results translate into a difference in binary fraction 
(control\,$-$\,planet-host) of $8.2\pm5.0$\% for the full subsample and of 
$12.5\pm5.9$\% for the redefined one. Although the relative errors on 
these results are quite
large due to the small number of available companions, both sample
definitions yield a positive difference with a statistical significance of
1.6-2.1$\sigma$. In physical terms, this positive difference means that planets 
(mainly giant ones) are less frequent in binaries with mean
semimajor axes between 35 and 250 AU than around single stars. In other words,
stellar duplicity seems to negatively impact planet occurrence in 
binaries with mean semimajor axes between 35 and 250 AU.

\begin{figure}[!ht]
\begin{center}
\plottwo{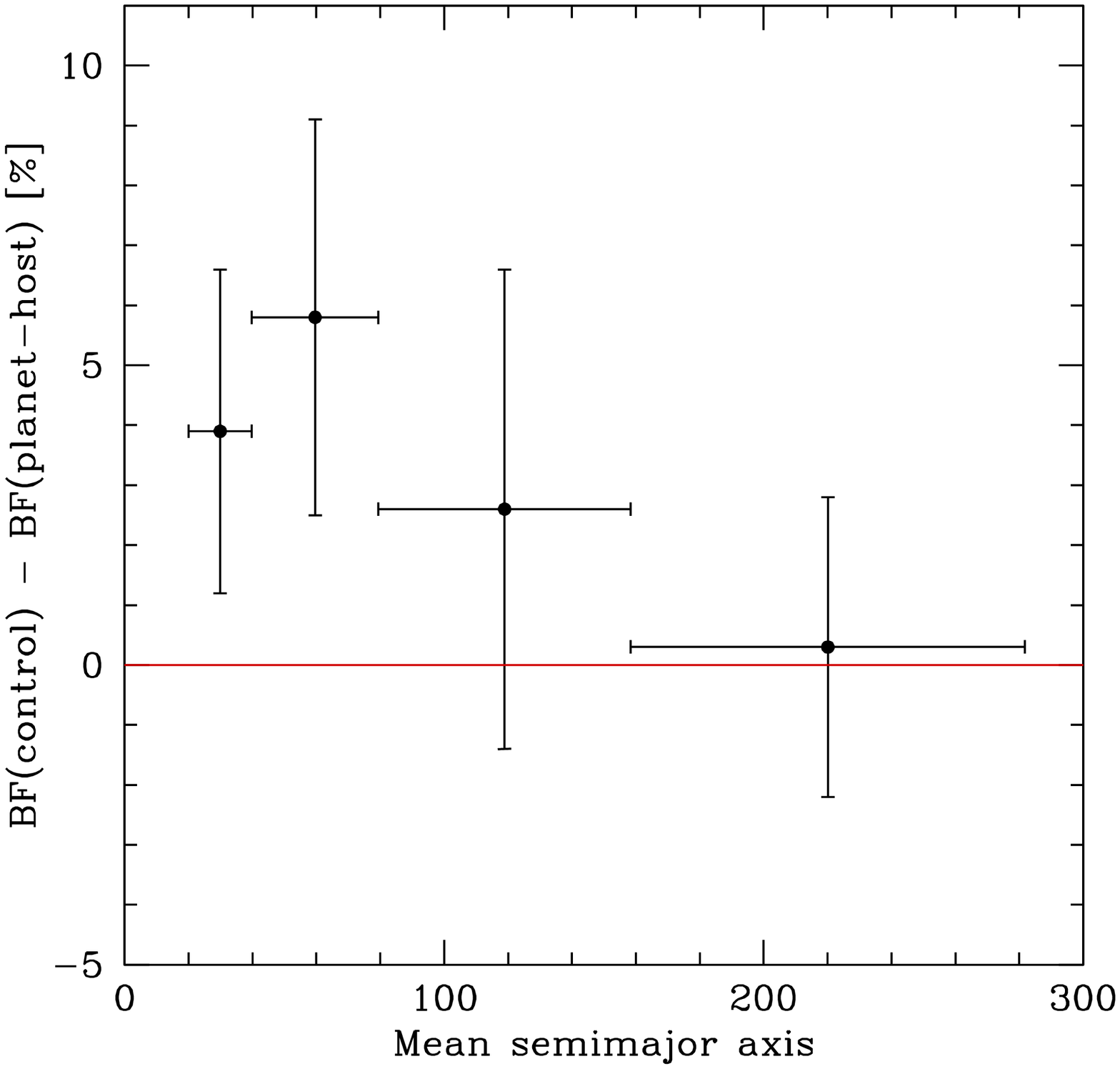}{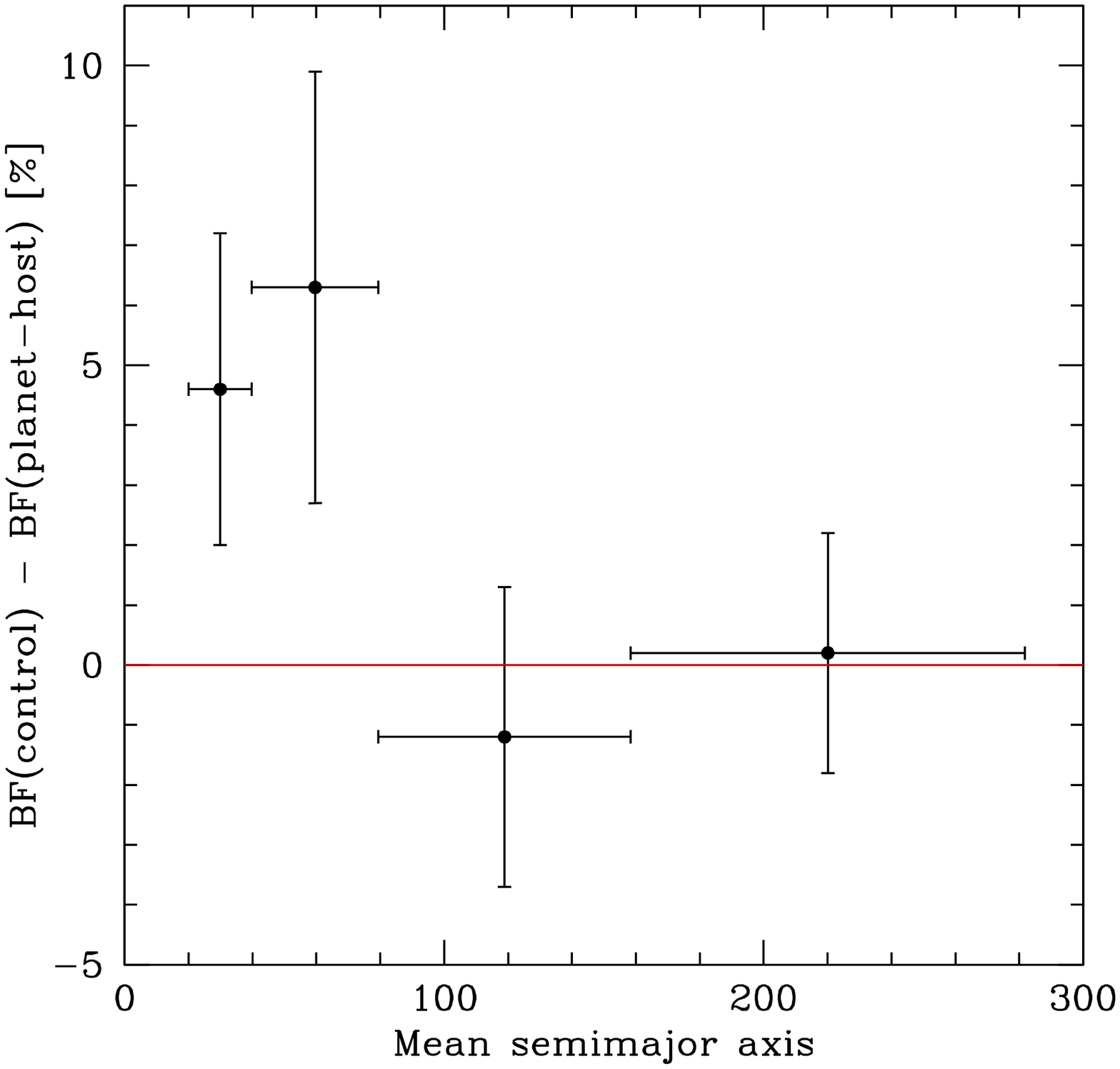}
\end{center}
\caption{Difference (in per cent) between the binary fraction of control stars
and the binary fraction of planet-host stars as a function of binary mean
semimajor axis. The left plot is based on the redefined subsamples, while the 
right plot is based on the full subsamples.}
\label{stat_res}
\end{figure}

To push the investigation a step further and to seek for a possible trend 
with mean semimajor axis, we computed the difference in binary fraction for
a few bins (equally spaced in logarithmic scale) between 20 and 280 AU. The results for
both the full and redefined subsamples are displayed on Fig.~\ref{stat_res}. These two
plots show that the difference in binary fraction seems not spread uniformly over
the semimajor axis range studied, but seems rather concentrated below $\sim$100
AU. Taken at face value, this result is very appealing as it might corroborate
the theoretical studies that predict a negative impact of stellar duplicity on
planet formation in binaries closer than $\sim$100 AU. Nonetheless, as 
just mentioned and as partly visible on Fig.~\ref{stat_res}, the small number
of true companions available for the statistics severely limits our analysis.
Larger samples will thus be needed to confirm the trends discussed here.

\section{Concluding Remarks and Future Prospects}

The preliminary statistical results presented here are quite encouraging and 
interesting as 
they might constitute the first observational evidence that circumstellar giant 
planets are less frequent in binaries closer than $\sim$100 AU than around
single stars. By adding about 70 stars to the statistics, the future
results from our complementary survey with PUEO will make a valuable 
contribution to the analysis and will improve the statistical significance of
the present results. It is worth recalling that the use of a well-defined control subsample
ensures that the apparent lower frequency of planets in binaries closer than
$\sim$100 AU is not due to selection effects against close binaries in 
Doppler planet searches. In this respect, the preliminary statistical analysis 
presented here goes beyond what has been done so far, as the former analyses 
\citep{Patience02,Raghavan06,Bonavita07} 
could not correct their results for these selection effects.

One point on which all the observational studies agree, is that 
if stellar duplicity impacts the formation and/or survival of circumstellar 
giant planets in some types of binaries, this effect is not easy to 
identify and to quantify in practice. This conclusion may result from 
practical limitations in the surveys (small sample sizes, difficulty to correct
for selection effects, practical impossibility to ensure that control stars are 
free from planets, ...), but it may alternatively have a more physical origin 
(not only binary semimajor axis but also eccentricity and mass ratio 
likely play a key role in determining the impact of stellar duplicity on planet
formation and evolution, dynamical evolution may significantly alter the
initial distributions and destroy the imprints of the formation process, ...). 
Recent Doppler searches for planets in spectroscopic binaries constitute
another avenue to study the impact of stellar duplicity on giant planet 
occurrence. Since these programs typically target binaries closer than $\sim$40
AU, their future results and conclusions will nicely complement those of 
the present imaging surveys. Our understanding of the impact of stellar 
duplicity on planet occurrence should thus significantly improve within 
the next years.

\acknowledgements 
A.~E. acknowledges support from the Swiss National Science Foundation through
a fellowship for prospective researcher. She also thanks the Soci\'et\'e Suisse
d'Astronomie et d'Astrophysique and the French Agence Nationale de la Recherche 
for financial support to attend the conference.


\end{document}